\documentclass[reprint, amsmath,amssymb, aps, showkeys]{revtex4-2}
\usepackage{graphicx}
\usepackage{dcolumn}
\usepackage{bm}
\usepackage{amsmath}
\usepackage{float}
\usepackage{subfigure}
\usepackage{booktabs}
\usepackage{CJKutf8}
\usepackage[colorlinks=true, linkcolor=blue, urlcolor=blue, citecolor=blue, linktoc=all]{hyperref}

\usepackage{tikz,xcolor,hyperref}
\definecolor{lime}{HTML}{A6CE39}
\DeclareRobustCommand{\orcidicon}{%
	\begin{tikzpicture}
	\draw[lime, fill=lime] (0,0) 
	circle [radius=0.16] 
	node[white] {{\fontfamily{qag}\selectfont \tiny ID}};
	\draw[white, fill=white] (-0.0625,0.095) 
	circle [radius=0.007];
	\end{tikzpicture}
	\hspace{-2mm}
}

\foreach \x in {A, ..., Z}{%
	\expandafter\xdef\csname orcid\x\endcsname{\noexpand\href{https://orcid.org/\csname orcidauthor\x\endcsname}{\noexpand\orcidicon}}
}

\begin{document}
\begin{CJK*}{UTF8}{gbsn}

\title{Subtraction of the confusion foreground and parameter uncertainty of resolvable galactic binaries on the networks of space-based gravitational-wave detectors}

\author{Jie Wu (吴洁) \orcidA{}}
\author{Jin Li (李瑾) \orcidB{}}
\email{cqujinli1983@cqu.edu.cn}
\affiliation{College of Physics, Chongqing University, Chongqing 401331, China}
\affiliation{Department of Physics and Chongqing Key Laboratory for Strongly Coupled Physics, Chongqing University, Chongqing 401331, China}

\begin{abstract}
There are tens of millions of compact binary systems in the Milky Way, called galactic binaries (GBs), most of which are unresolved, and the Gravitational waves (GWs) emitted overlap to form confusion foreground. 
By simulating such confusion foreground, we have studied how LISA, Taiji and TianQin, including their alternative orbital configurations, subtract resolvable GBs when they combine as some networks.
The results of our research indicate that the order of detected number for a single detector from high to low is: Taiji-m, Taiji-p (c), LISA, TianQin I, TianQin II.
For detector combinations on the network, the confusion foreground is effectively reduced as the number of detectors grows, and the optimal combinations with different numbers are: Taiji-m, LISA+Taiji-m, LISA+Taiji-m+TianQin I, and LISA+Taiji-m+TianQin I+II.
The sensitivity curve is optimized as the number of detectors increases, which renders it possible to detect other gravitational wave sources more precisely and decrease the resolvable GBs parameter uncertainty.
Based on this, we discuss the parameter uncertainty of resolvable GBs detected by the combinations above and find that GW detection can promote electromagnetic (EM) detection.
On the contrary, we discovered that by utilizing EM detection, determining the inclination angle can reduce the uncertainty of GW strain amplitude by $\sim$93\%, and determining the sky position can reduce the uncertainty of the phase by $\sim$30\%, further strengthening the connection between GW detection and EM detection, and contributing to the research of Multi-messenger astronomy.
\end{abstract}

\maketitle
\end{CJK*}

\section{Introduction}
Since LIGO detected the first GW event from a binary black hole merger (GW150914) in 2015~\cite{GW150914}, a series of ground-based GW detectors, such as Advanced LIGO~\cite{aLIGO},
Advanced Virgo~\cite{aVIRGO,VIRGO} and KAGRA~\cite{KAGRA1,KAGRA2}, have been built around the world, opening the window for detecting GW.
However, due to the limitation of the interferometer arm length, the observation window of the ground-based GW detector is in the high-frequency band from 1 Hz to kHz, and the low-frequency GW signal below 1 Hz cannot be effectively detected.
Therefore, constructing an interferometer with an arm length in order of one million kilometers in space is an ideal solution for detecting low-frequency GW.

The mission proposed by European Space Agency to detect GW in the low-frequency band named Laser Interferometer Space Antenna (LISA) is scheduled to be launched around the 2030s~\cite{LISA}.
At the same time, the Taiji mission proposed by the Chinese Academy of Sciences to construct a space-based GW observatory similar to LISA, which consists of a triangle of three spacecraft (S/C) orbiting the sun linked by laser interferometers, will be in operation~\cite{Taiji}.
Another Chinese mission, TianQin, being different from LISA and Taiji, consists of three identical drag-free controlled S/C in high Earth orbits~\cite{TianQin}.
LISA, Taiji, and TianQin are all sensitive to the milli-Hertz frequency band ($\sim10^{-4}-10^{-1}$ Hz).
Indeed, there are some notional space-based GW missions in the frequency band spanning from milli-Hertz to Hertz.
One such mission is the Deci-Hertz Interferometer Gravitational Wave Observatory (DECIGO) proposed by Japan~\cite{DECIGO}.
Additionally, there are missions like the Advanced Laser Interferometer Antenna (ALIA) and the Big Bang Observer (BBO), which are considered potential follow-on missions to LISA, operating in similar frequency band~\cite{Beyond_LISA}. 
These missions aim to explore different frequency bands and provide complementary insights into the universe.

Compared with the Hertz frequency band, there are a large variety of GW sources in the milli-Hertz frequency band sensitive to the space-based GW detectors. These sources are expected to carry a large amount of information about galaxy formation, galactic nuclei, the Milky Way, and the early universe~\cite{source1,source2}, including massive black hole binaries (MBHB)~\cite{MBHB}, extreme/intermediate mass ratio inspirals (EMRIs/IMRIs)~\cite{EMRI}, compact binaries
in the Milk Way~\cite{compact_binaries1,compact_binaries2} and stochastic gravitational-wave backgrounds (SGWBs)~\cite{SGWB1,SGWB2}.

According to current astrophysical models and observations, there are a large number of GBs in our Milky Way, whose orbital period is less than a few hours, and the frequency band of emitted GW is from 0.1 mHz to 10 mHz~\cite{astrophysical_model1,astrophysical_model2}.
Considering the sensitivity of the space-based GW detectors, the GWs emitted by tens of millions of GBs will enter the observation frequency band at the same time, overlapping to form the galactic foreground~\cite{astrophysical_model3,astrophysical_model4,astrophysical_model5}.
Except for a small percentage of high signal-to-noise ratio (SNR) GBs known as resolvable GBs, the majority of them are unresolved, resulting in an effective noise called confusion foreground or confusion noise.~\cite{on_networks,Characterization_LISA}.
In the frequency range of 0.5$\sim$3 mHz, the confusion foreground will be greater than the instrument noise, affecting the observation of other GW sources and creating a bump on the sensitivity curve.
While the unresolved GBs constitute the confusion foreground and have a negative impact on the observation of other GW sources, the resolvable GBs are conducive to researching the evolution and distribution of GBs in our Milky Way, which is also one of the main science objectives of the space-based GW detectors~\cite{science_objectives1,science_objectives2}.

Since the proposal of LISA, extensive research has been conducted on the confusion foreground from GBs~\cite{proposal_LISA}.
Cornish et al. first presented an analytical fitting function for LISA's confusion foreground at the 11th International LISA Symposium, which subsequent studies have built upon~\cite{LISA_GBs}.
In Ref.~\cite{TaijiGBs}, Liu et al. estimated the confusion foreground for Taiji by subtracting resolvable GBs and compared it with LISA.
Huang et al., in Ref.~\cite{TianQinGBs}, investigated the case of TianQin considering different configurations and its joint observation with LISA.
Unlike LISA, where the confusion foreground is primarily caused by double white dwarfs, for GW space-based detectors like BBO, the confusion foreground is predominantly composed of neutron-star (NS) binary.
In Ref.~\cite{BBO_NS}, Cutler et al. discussed the impact of the NS binary subtraction problem and baseline design.
Yagi et al., in Ref.~\cite{BBO_DECIGO}, studied the subtraction of NS binaries for DECIGO and BBO with different configurations.
Furthermore, in addition to increasing observation time and improving the sensitivity of the GW detector, Refs.~\cite{network,network2} reported that the networks of the GW detector can also effectively identify more resolvable GBs and subtract the confusion foreground.

In this paper, we simulate subtracting confusion foreground using different combinations between LISA, Taiji, and TianQin on the network, including their alternative orbital configurations to determine the best combination on the network, and draw the sensitivity curve to calculate the SNR and parameter uncertainty of detected resolvable GBs, thus discussing the Multi-messenger astronomy combined with EM detection.

This paper is organized as follows.
In Sec.~\ref{sec:signals_and_detectors}, we introduce the GW signal model used to simulate GBs, the response of different space-based GW detectors to GW, as well as their instrument noise, sensitivity, and the alternative orbit configurations.
In Sec.~\ref{sec:Methodology}, we use the population model to construct the GBs signal, subtracting the resolvable GBs by the iterative procedure to estimate the confusion foreground, and calculating the parameters of the resolvable GBs.
In Sec.~\ref{sec:Results}, we present the subtraction of the confusion foreground by different combinations on the network, analyze the factors responsible for them, and plot the full sensitivity curves containing the confusion foreground.
Finally, we summarize our results in Sec.~\ref{sce:Summary}.

\section{GW signals and detectors}\label{sec:signals_and_detectors}
\subsection{GW signals from GBs}\label{sec:signals_from_GBs}
Considering that GBs have a few hours of orbital period and emit GW frequencies in milli-Hertz, they are in the very beginning phase of inspiral, millions of years before the merger~\cite{astrophysical_model1,GBhour}.
Therefore, the orbital period evolves slowly and the GWs emitted by GBs can be fully regarded as quasi-sinusoidal signals (quasi-monochromatic sources).
For the GW signal, we can use a very simple model in which the phase is decomposed in a Taylor series, and consequently, the time domain waveform of a GB can be written as~\cite{Maggiore}:
\begin{equation}\label{Eq:waveform}
    \begin{aligned}
        & h_{+}(t)=\mathcal{A}(1+\cos\iota^{2})\cos\Phi(t) \\
        &h_{\times}(t)=2\mathcal{A}\cos\iota\sin\Phi(t)
    \end{aligned}
\end{equation}
with\begin{equation}
     \Phi(t)=\phi_0+2\pi f_0t+\pi\dot{f}_0t^2+ \Phi_{D}(t)
\end{equation}
where $\mathcal{A}$ is the GW strain amplitude, $\iota$ is the inclination angle, $\Phi(t)$ is the orbital phase , $\Phi_D(t)$ is the Doppler phase, $\phi_0$ is the initial phase, $f_0$ and $\dot{f}_0$ is the frequency and the derivative of the frequency of GW.
The frequency variation, also known as the frequency derivative, can be expressed with the equation described in Ref.~\cite{frequency_derivative}:
\begin{equation}\label{Eq:frequency_derivative}
    \dot{f}_0=\frac{96}{5}\left(\frac{G\mathcal{M}}{c^3}\right)^{5/3}f_0^{11/3}
\end{equation}
where $\mathcal{M}=(m_1m_2)^{3/5}/(m_1+m_2)^{1/5}$ is the chirp mass, $G$ and $c$ are the gravitational constant and the speed of light.
By substituting frequency $f_0\sim10^{-3}$ into equation 3, we can roughly calculate the derivative of frequency $\dot{f}_0\sim10^{-19}$, indicating that the derivative of frequency is much lower in magnitude than that of frequency, which is also why we consider GWs as  quasi-sinusoidal signals.
Therefore, we neglect higher-order phase terms as they contribute minimally to the waveform and have little impact on confusion foreground.
Additionally, we assume that the GBs are in circular orbits and ignore the influence of the third perturbation body.~\cite{circular_orbit1,circular_orbit2}.

For the space-based GW detector, the periodic motion around the Sun will produce the Doppler phase, which is given by~\cite{TianQin_orbit}:
\begin{equation}
    \Phi_{D}(t) = 2\pi f_0(R/c)\cos\beta\cos(2\pi f_mt-\lambda )
\end{equation}
where $R$ = 1 A.U. is the distance between the Sun and the Earth, $f_m$ = 1/year is the Geocentric orbit modulation frequency and ($\lambda,\beta$) are the Ecliptic coordinates of the GW source.

\subsection{Detector’s response and noise}
\begin{figure}[ht]
    \begin{minipage}{\columnwidth}
        \centering
        \includegraphics[width=0.9\textwidth,
        trim=0 0 0 0,clip]{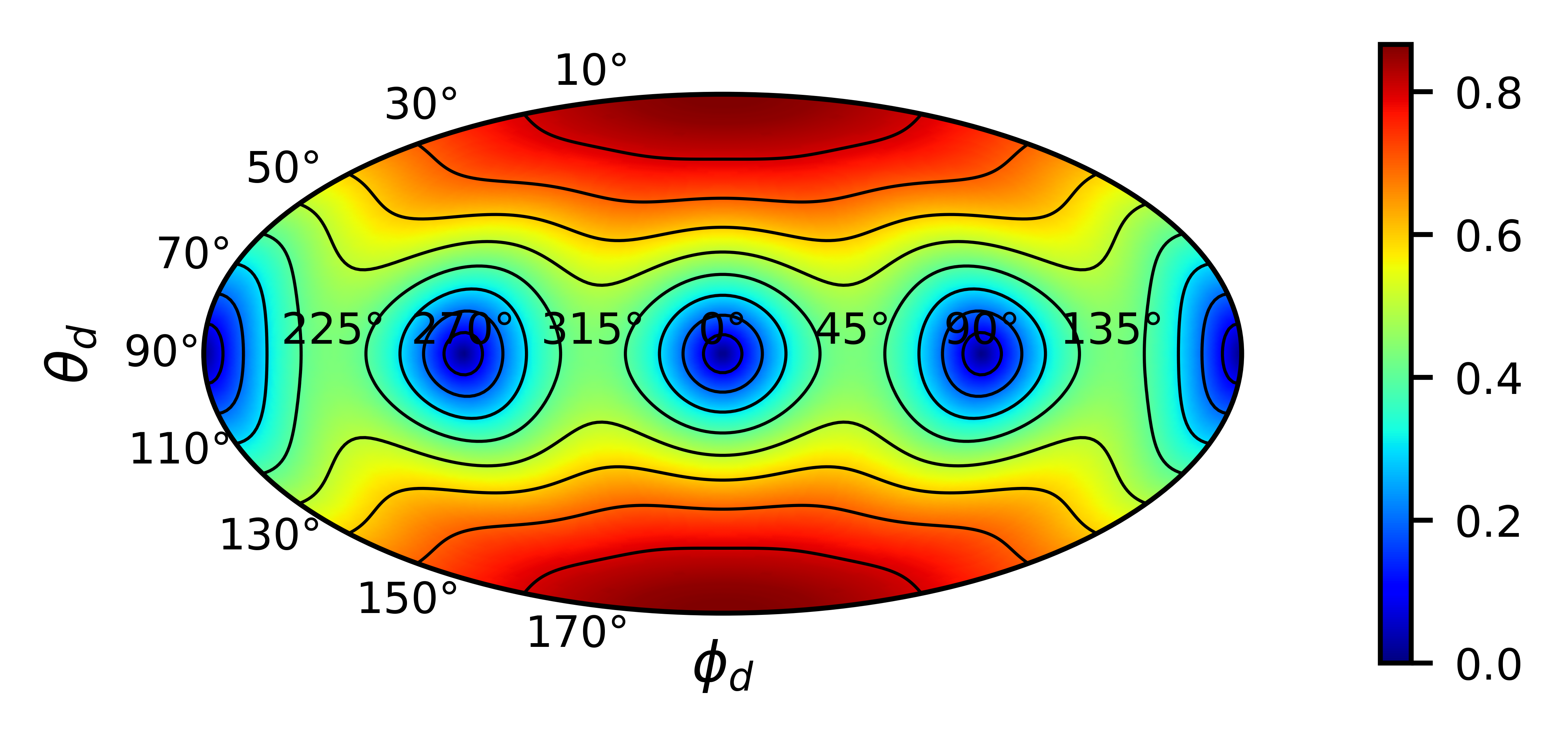}
        \caption{The combined tensor mode response function $F$ in the detector coordinate frame.}\label{FIG:response}
    \end{minipage}
\end{figure}

For the space-based GW detector, the GW strain recorded by the detector can be described as the linear combination of two GW polarizations~\cite{strain_in_detector}:
\begin{equation}
    h(t)=F^+(t)h_+(t)+F^\times(t)h_\times(t)
\end{equation}
where $F^+$ and $F^\times$ are the antenna pattern functions.
In the low-frequency limit, the antenna pattern functions in the detector’s coordinate frame can be expressed as\cite{response_function}:
\begin{widetext}
    \begin{equation}
        \begin{aligned}
            F^{+} & = -\frac{\sin\gamma}{2}[(1+\cos^{2}\theta_{d})\sin2\phi_{d}\cos2\psi_s+2\cos\theta_{d}\cos2\phi_{d}\sin2\psi_s] \\
            F^{\times} & = -\frac{\sin\gamma}{2}[-(1+\cos^{2}\theta_{d})\sin2\phi_{d}\sin2\psi_s+2\cos\theta_{d}\cos2\phi_{d}\cos2\psi_s] 
        \end{aligned}
    \end{equation}
\end{widetext}
where $\gamma=\pi/3$ is the angle between the two arms of the detector, ($\phi_d$,$\theta_d$) are the coordinates of the location of the GW source in the
detector coordinate frame and $\psi_s$ is the polarization angle.
The transformation between detector coordinates ($\phi_d$,$\theta_d$) and Ecliptic coordinates ($\lambda$,$\beta$) can be found in Appendix~\ref{App:Coordinate_transformation}.
To explore the response of the detector to GWs in different positions, we introduce the combined tensor mode response function:
\begin{equation}
    F=\sqrt{|F^+|^2+|F^\times|^2}
\end{equation}
The results in the detector coordinate frame are shown in FIG.~\ref{FIG:response}. 
It can be seen that the position perpendicular to the constellation plane has the highest response, implying that different orientations will affect detection capacity in the same configuration.

Besides, the noise of the detector is another element that influences detection ability.
In this paper, we focus solely on the impact of instrument noise composed of acceleration noise and displacement noise when subtracting confusion foreground.
Therefore, an analytical model of the detector's sensitivity curve $S_n(f)$ can be constructed from the sky average response function and instrument noise.
For LISA~\cite{sensitivity_curve} and Taiji~\cite{source1,Taiji_science}, the sensitivity curve can be expressed as follows:
 \begin{equation}
    \begin{aligned}\label{Eq:sen_curve1}
        S_n(f)& =\frac{10}{3L^{2}}\left[P_{\mathrm{dp}}+2(1+\cos^{2}(f/f_{*}))\frac{P_{\mathrm{acc}}}{(2\pi f)^{4}}\right]  \\
        &\times\left[1+0.6\left(\frac{f}{f_*}\right)^2\right]
    \end{aligned}
\end{equation}
with
\begin{equation}
    P_{\mathrm{dp}} =S_x\left[1+\left(\frac{2\mathrm{mHz}}{f}\right)^{4}\right] 
\end{equation}
\begin{equation}
    P_\mathrm{acc} =S_a\left[1+\left(\frac{0.4\operatorname{mHz}}{f}\right)^{2}\right]\left[1+\left(\frac{f}{8\operatorname{mHz}}\right)^{4}\right] 
\end{equation}
For TianQin~\cite{TianQin_orbit,TianQinGBs}, the sensitivity curve can be written in the form of:
\begin{equation}
    \begin{aligned}\label{Eq:sen_curve2}
        S_{n}(f)& =\frac{1}{L^2}\left[\frac{4S_a}{(2\pi f)^4}\left(1+\frac{0.4\operatorname{mHz}}{f}\right)+S_x\right] \\
        &\times\left[1+0.6\left(\frac{f}{f_*}\right)^2\right]
    \end{aligned}
\end{equation}
where $f_*=c/(2\pi L)$ is the transfer frequency, $c$ is the speed of light, $L$ is the arm length, $S_a$ is acceleration noise and $S_x$ is displacement measurement noise, all of which are given in TABLE~\ref{Tab:noise}.

\begin{table}[ht]
\centering
\renewcommand{\arraystretch}{1.5}
\caption{Noise and arm length of different detector}\label{Tab:noise}
\begin{tabular*}{\columnwidth}{@{\extracolsep{\fill}}cccc@{}}
\hline
 Detector & $S_a/\mathrm{m}^{2}\mathrm{s}^{-4}\mathrm{Hz}^{-1}$ & $S_x/\mathrm{m}^{2}\mathrm{Hz}^{-1}$ & $L/\mathrm{m}$ \\
\hline
LISA & $9\times10^{-30}$ & $225\times10^{-24}$ & $2.5\times10^{9}$\\
Taiji & $9\times10^{-30}$ & $64\times10^{-24}$ &$3\times10^{9}$  \\
TianQin & $1\times10^{-30}$ & $1\times10^{-24}$ &$\sqrt{3}\times10^{8}$ \\ 
\hline
\end{tabular*}
\end{table}

\subsection{Alternative orbital configurations}
\begin{figure}[ht]
    \begin{minipage}{\columnwidth}
        \centering
        \includegraphics[width=0.95\textwidth,
        trim=0 0 0 0,clip]{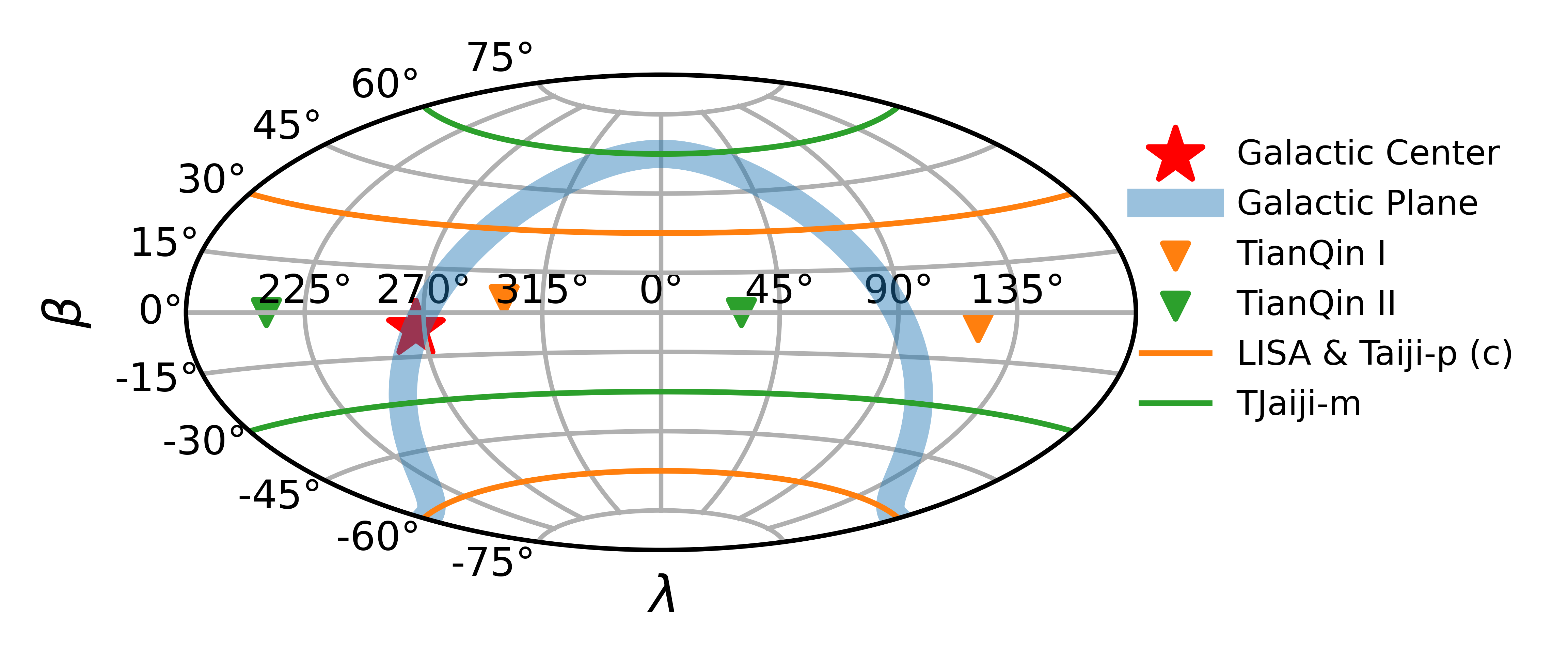}
        \caption{The variation of the normal direction of the detector constellation plane during the observation period.}\label{FIG:normal_direction}
    \end{minipage}
\end{figure}

\begin{figure*}[ht]
    \begin{minipage}{\textwidth}
        \centering
        \includegraphics[width=0.9\textwidth,
        trim=0 0 0 0,clip]{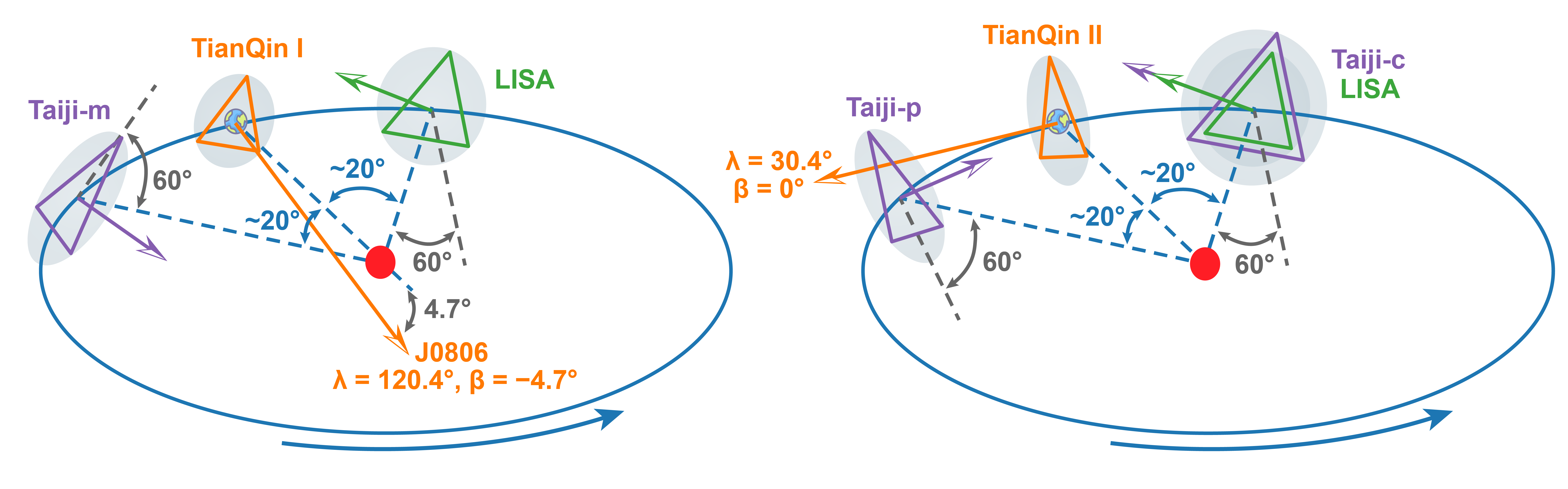}
        \caption{The alternative orbital configurations of LISA, Taiji-p, Taiji-c, Taiji-m, TianQin I and TianQin II. Note that this is is only a schematic figure of alternative orbital configurations, not the actual scale. The solid arrow on the constellation plane represents the normal direction of the detector constellation plane, as shown in FIG.~\ref{FIG:coordinate} of Appendix~\ref{App:Coordinate_transformation} in different coordinate frames.}\label{FIG:orbit}
    \end{minipage}
\end{figure*}

LISA, Taiji, and TianQin are all scheduled to launch a triangular constellation composed of three S/C. 
The difference is that LISA and Taiji apply heliocentric orbits, whereas TianQin applies geocentric orbits. 
There are multiple orbital configurations to be chosen, as detailed in  FIG.~\ref{FIG:normal_direction}, FIG.~\ref{FIG:orbit} and TABLE~\ref{Tab:orbit}.
\begin{table}[ht]
\centering
\renewcommand{\arraystretch}{1.3}
\caption{Alternative orbital configurations}\label{Tab:orbit}
\begin{tabular*}{\columnwidth}{@{\extracolsep{\fill}}crr@{}}
\hline
           & Inclination & Leading angle \\ \cline{2-3} 
LISA       & $60^\circ$  & $-20^\circ$    \\
Taiji-p    & $60^\circ$  & $20^\circ$     \\
Taiji-c    & $60^\circ$  & $-20^\circ$    \\
Taiji-m    & $-60^\circ$  & $20^\circ$    \\ \hline
           & $\lambda$         & $\beta$       \\ \cline{2-3} 
TianQin  I & $120.4^\circ$ & $-4.7^\circ$   \\
TianQin II & $30.4^\circ$  & $0^\circ$ \\ \hline
\end{tabular*}
\end{table}

LISA includes three S/C forming a $2.5\times10^6$ km triangle trailing the Earth by $20^\circ$ on the Heliocentric orbit and the constellation plane has a $60^\circ$ inclination to the Ecliptic plane as shown in FIG.~\ref{FIG:normal_direction} and FIG.~\ref{FIG:orbit}.
Meanwhile, Taiji expects to use a LISA-like orbital configuration with a $3\times10^6$ km arm length and three different orbital configuration options available~\cite{TaijiGBs,Alternative_Taiji1,Alternative_Taiji2}.
The first configuration is called Taiji-p, which has the same inclination angle as LISA but is $20^\circ$ ahead of Earth. The second configuration is exactly the same as LISA, called Taiji-c. These two configurations are shown on the right side of FIG.~\ref{FIG:orbit}. The third configuration named Taiji-m has an inclination of $-60^\circ$ to the Ecliptic plane and a leading angle of $20^\circ$ to the Earth, as shown on the left side of FIG.~\ref{FIG:orbit}.

Unlike LISA and Taiji, TianQin uses a Geocentric orbit with a $\sqrt{3}\times10^5$ km arm length, hence the normal direction of the constellation plane will remain unchanged, pointing in the same direction.~\cite{TianQin,TianQinGBs}.
The two orbital configurations of TianQin are the different orientations of the normal directions of the constellation plane.
The normal direction of TianQin I points towards the tentative reference source RX J0806.3+1527 (pointing towards $\lambda = 120.4^\circ,\beta = -4.7^\circ$), while the normal direction of TianQin II perpendicularly (pointing towards $\lambda = 30.4^\circ,\beta = 0^\circ$), which is shown in FIG.~\ref{FIG:normal_direction} and FIG.~\ref{FIG:orbit}.

The observation time varies with different orbital configuration. 
LISA and Taiji are both year-round observation schemes, and any of Taiji's three alternative orbital configurations will not operate simultaneously.
Different from the former, TianQin follows the “three months on + three months off” observation scheme, and TianQin I and TianQin II can operate simultaneously to fill the data gaps of each other~\cite{TianQinGBs}, which will be considered in the subtraction methodology in Sec.~\ref{sec:Subtraction}.

\section{Methodology}\label{sec:Methodology}
\subsection{Data analysis}
The SNR $\rho$ of a GB source, which play an important role for judging the resolvable sources, can be defined as:
\begin{equation}\label{Eq:SNR}
    \rho^2=(h|h)
\end{equation}
where the inner product $(\cdot|\cdot)$ is a generalisation of the time-domain correlation product and is conventionally defined as~\cite{strain_in_detector,inner_product}:
\begin{equation}
    \begin{aligned}\label{Eq:inner_product}
        (a|b)& =4\int_0^\infty\mathrm{d}f\ \frac{\tilde{a}^*(f)\tilde{b}(f)}{S_n(f)}  \\
        &\simeq\frac{2}{S_n(f_0)}\int_0^{T_{obs}}\mathrm{d}t\ a(t)b(t)
    \end{aligned}
\end{equation}
where $\tilde{a}(f)$ and $\tilde{b}(f)$ are the Fourier transformations of $a(t)$ and $b(t)$, $S_n(f)$ is the sensitivity curve defined by Eq.~\ref{Eq:sen_curve1} and Eq.~\ref{Eq:sen_curve2}, $T_{obs}$ is the observation duration. 
Note that the second line of Eq.~\ref{Eq:inner_product} only holds when calculating a quasi-sinusoidal signal (quasi-monochromatic source) that have an almost constant noise PSD and it can be seen that the SNR increases while the observation duration increases.
A quasi-sinusoidal signal like GB can be represented in the spectrum using the Dirac Delta function, thus the signal is plotted as a point with amplitude in the spectrum. Therefore, the SNR of GB in the Eq.~\ref{Eq:SNR} can be roughly calculated as follows, which is obtained by evaluating the SNR integral~\cite{sensitivity_curve}:
\begin{equation}\label{Eq:optimal_SNR}
    \overline{\rho^2} =\frac{16}{5} \frac{\mathcal{A}^2T_{obs}}{S_n(f_0)} 
\end{equation}
where $\mathcal{A}$ is the GW strain amplitude.
Using Eq.~\ref{Eq:optimal_SNR} can calculate SNR more quickly than using Eq.~\ref{Eq:SNR}, and in the processing steps of Sec.~\ref{sec:Subtraction}, we use Eq.~\ref{Eq:optimal_SNR} to quickly calculate and filter optimal resolvable GBs.

Usually, the GB with the SNR greater than 7 ($\rho>7$) is defined as the resolvable GB~\cite{Characterization_LISA,LISA_GBs} and we can analyze the uncertainties of the resolvable GB using Fisher information matrix (FIM), which is defined as:
\begin{equation}\label{Eq:FIM}
    \Gamma_{ij}=\left(\frac{\partial h}{\partial\xi_i}\middle|\frac{\partial h}{\partial\xi_j}\right)
\end{equation}
where $\xi_i$ represents the parameter of GB. For high SNR signals ($\rho\gg 1$), the variance-covariance matrix obtained from the inverse of FIM, $\Sigma=\Gamma^{-1}$, where the diagonal element represents the variance (or mean squared error) of each parameter, and the off-diagonal element represents the covariance (or correlation) between the parameters~\cite{TianQinGBs,inner_product}.
Therefore, the uncertainty of each parameter can be written as:
\begin{equation}
    \Delta\xi_i=\sqrt{\Sigma_{ii}}
\end{equation}
Compared to the uncertainty of coordinates, the uncertainty of sky position is more commonly used, which can be obtained by combining the uncertainty of both coordinates~\cite{strain_in_detector}:
\begin{equation}
    \Delta\Omega=2\pi\big|\sin\beta\big|\sqrt{\Sigma_{\beta\beta}\Sigma_{\lambda\lambda}-\Sigma_{\beta\lambda}^2}
\end{equation}

When calculating FIM in Eq.~\ref{Eq:FIM}, use the following numerical differentiation approximation~\cite{TianQinGBs,inner_product}:
\begin{equation}
    \frac{\partial h}{\partial\xi_{i}}\approx\frac{h(t,\xi_{i}+\delta\xi_{i})-h(t,\xi_{i}-\delta\xi_{i})}{2\delta\xi_{i}} 
\end{equation}

When considering network detection by multiple independent detectors, the total SNR and FIM can be obtained by the sum of the inner products calculated by each detector, which can be written as~\cite{TianQinGBs}:
\begin{equation}
    \begin{aligned}\label{Eq:net_SNR_FIM}
        \rho_\mathrm{net}^2=\sum_{k}{\rho_k^2}=\sum_{k}(h_{k}|h_{k})\\
        \Gamma_\mathrm{net}=\sum_{k}{\Gamma_k}=\sum_k\left(\frac{\partial h_k}{\partial\xi_i}\middle|\frac{\partial h_k}{\partial\xi_j}\right) 
    \end{aligned}
\end{equation}
where $k$ represents different independent detectors.
From Eq.~\ref{Eq:net_SNR_FIM}, the sensitivity in the network can be obtained, whose reciprocal is the sum of the reciprocal sensitivities of each detector, which can be expressed as follows:
\begin{equation}\label{Eq:net_sensitivity}
    S_{\mathrm{net}}^{-1}=\sum_{k}S_{k}^{-1}
\end{equation}

\subsection{Subtraction of the confusion foreground }\label{sec:Subtraction}
For population simulation of GBs, we used the population datasets from the first “new” LISA Data Challenge (LDC), codenamed \texttt{Radler}, which contains approximately 30 million GB sources in the milli-Hertz band~\cite{astrophysical_model3,LDC}.
For the convenience of data processing, we select 1\% of the GBs in \texttt{Radler} ($3\times10^5$ GBs) and multiply them to achieve the same amplitude level as the actual situation to generate the galactic foreground.
The number of $3\times10^5$ GBs is sufficient to include the same parameter distribution $3\times10^7$ GBs in \texttt{Radler}, and the number of the resolvable GBs should be 1\% of that in \texttt{Radler}.
Notice that although the multiplication operation was performed during the generation of the galactic foreground, which would increase the amplitude of a single signal, the smoothed spectrum is used in subsequent processing to obtain the same amplitude as \texttt{Radler} without affecting the calculated SNR.

\begin{figure*}[ht]
    \begin{minipage}{\textwidth}
        \centering
        \includegraphics[width=0.95\textwidth,
        trim=0 0 0 0,clip]{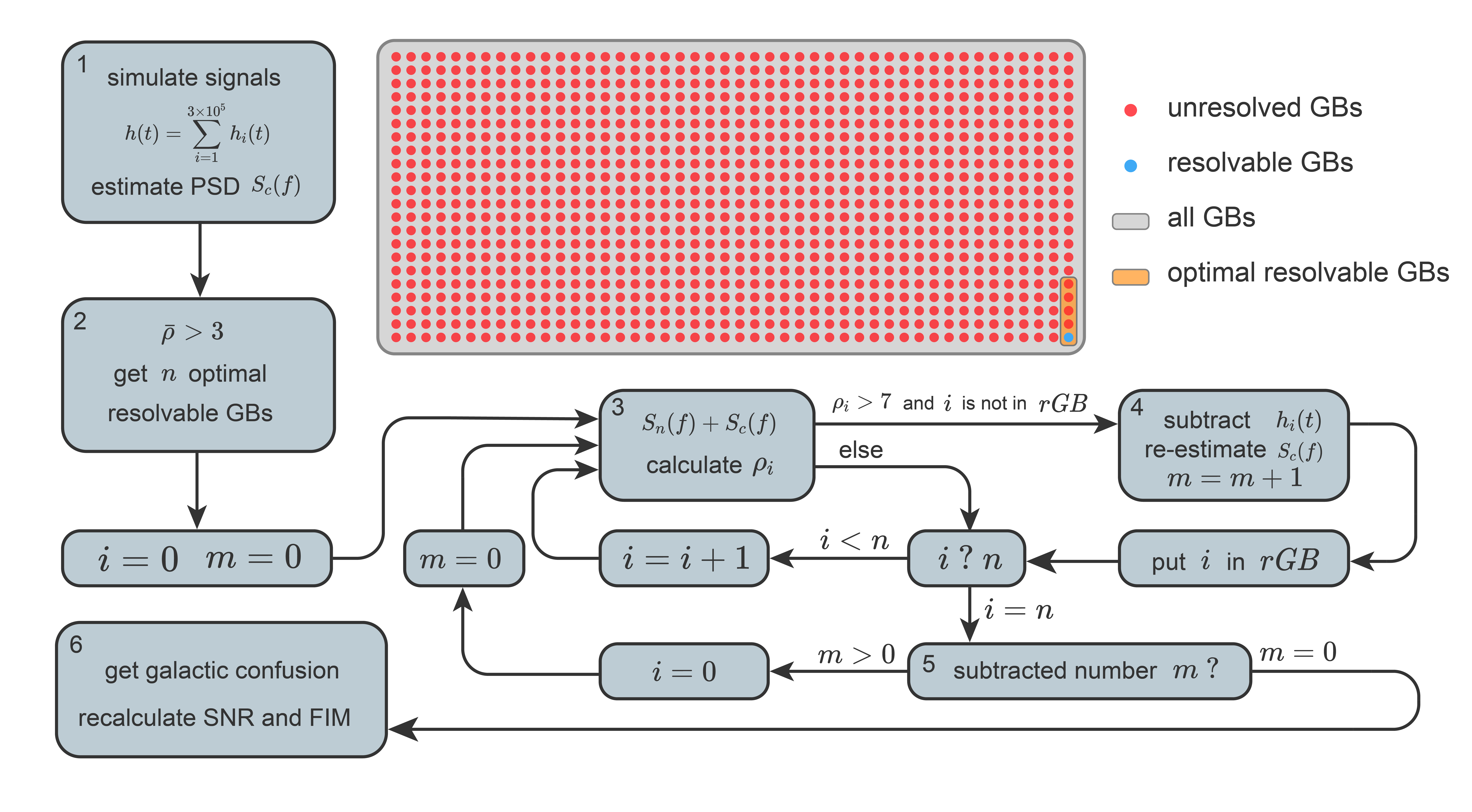}
        \caption{The flow chart of subtracting the confusion foreground and the proportion diagram of different GBs. The flowchart provides the basic steps for subtracting confusion foreground in Sec.~\ref{sec:Subtraction}. The proportion diagram shows the proportion of different GBs, indicating that the optimal resolvable GBs can filter out most unresolved GBs.}\label{FIG:subtration}
    \end{minipage}
\end{figure*}

The basic steps for subtracting confusion foreground are shown in FIG.~\ref{FIG:subtration}, which can be summarized as follows\cite{Characterization_LISA,TaijiGBs,TianQinGBs}:
\begin{enumerate}

    \item\label{step1} Simulate the superposition $h(t)$ of $3\times10^5$ GBs in the time domain and then calculate the power spectrum density (PSD) of the galactic foreground. Run the median on the PSD to estimate the confusion foreground $S_c(f)$.
    
    \item\label{step2} Roughly calculate the optimal SNR $\overline{\rho}$ under the sensitivity curve of instrument noise $S_n(f)$ using Eq.~\ref{Eq:optimal_SNR}, and consider GBs with an optimal SNR greater than 3 ($\overline{\rho}>3$) as optimal resolvable GBs, which can quickly filter out 99.6\% of unresolved GBs.
    
    \item\label{step3} For the $i$th optimal resolvable GB, the sensitivity curve is formed by adding instrument noise and confusion foreground ($S_n(f)+S_c(f)$), and the SNR $\rho_i$ is calculated using Eq.~\ref{Eq:SNR} and Eq.~\ref{Eq:inner_product}. If the SNR is less than 7 ($\rho_i<7$), skip and repeat the method to calculate the SNR of the ($i$+1)th optimal resolvable GB. If the SNR is greater than 7 ($\rho_i\ge7$), the GB is resolvable, and then continue with the next subtraction step.
    
    \item\label{step4} Subtract the $i$th GB signal in the time domain ($h(t)-h_i(t)$) and use the method in Step~\ref{step1} to re-estimate the subtracted galactic confusion. Repeat Steps~\ref{step3} and~\ref{step4}, continuously subtracting resolvable GBs and re-estimating galactic confusion until all optimal resolvable GBs are calculated.
    
    \item\label{step5} Repeat Steps~\ref{step3},~\ref{step4} and~\ref{step5} in the remaining optimal resolvable GBs until the subtracted GB is 0, indicating the galactic confusion composed of unresolved GBs.

    \item\label{step6} Recalculate the SNR and FIM of the resolvable GBs using the final subtracted galactic confusion.
\end{enumerate}
In the above steps, it is assumed that the resolvable GB can be subtracted perfectly without residual error, which will not be achievable in practice, and the subtraction error should be considered~\cite{subtraction_error1,subtraction_error2}.
When generating the time-domain galactic foreground signal, we set the Earth in the Vernal Equinox as zero time ($t=0$), and conduct observation simulation at different times ($T_{obs}=\{0.5,1,2,4\}\mathrm{years}$) to subtract the galactic confusion using the above basic steps. Considering the observation on the networks, we use the method of Eq.~\ref{Eq:net_SNR_FIM} to calculate the SNR and FIM, and get the results on different networks.

\section{Results}\label{sec:Results}
\subsection{Resolvable GBs}\label{sec:Resolvable_GBs}
\begin{figure*}[ht]
    \begin{minipage}{\textwidth}
        \centering
        \includegraphics[width=0.95\textwidth,
        trim=0 5 0 0,clip]{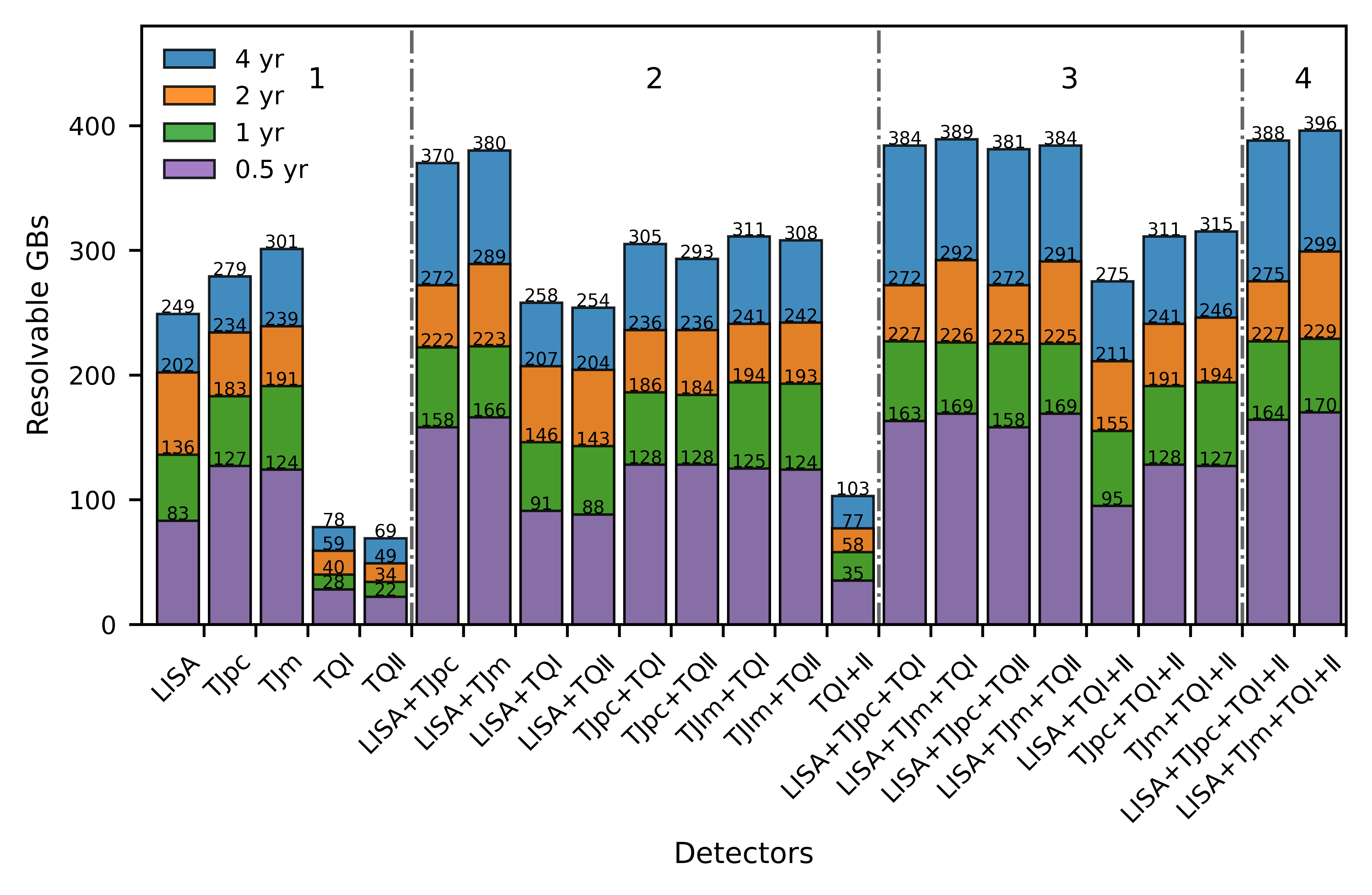}
        \caption{Quantitative statistics of resolvable GBs on different networks and different observation times.To make annotations more concise, \textit{TJ} represents Taiji and \textit{TQ} represents TianQin, which is also represented in FIG.~\ref{FIG:FIM}, FIG.~\ref{FIG:sensitivity} and TABLE~\ref{Tab:fit}. The results of Taiji-p and Taiji-c are the same and thus we use \textit{TJpc} to represent both. The numbers on the bars represent the number of resolvable GBs at different observation times ($T_{obs}=\{0.5,1,2,4\}\mathrm{years}$). Note that we only used $3\times10^{5}$ GBs, making the results approximately 1\% of the actual situation.}\label{FIG:RGB}
    \end{minipage}
\end{figure*}

Using the method in Sec.~\ref{sec:Subtraction}, we simulated and calculated the number of resolvable GBs detected on different detectors and their networks at different observation times, as shown in FIG.~\ref{FIG:RGB}.

Apparently, FIG.~\ref{FIG:RGB} illustrates that as the observation time increases, the number of resolvable GBs also increases due to Eq.~\ref{Eq:inner_product} and Eq.~\ref{Eq:optimal_SNR}.

Given the observation time, for a single detector, the number of resolvable GBs detected in descending order is: Taiji-m, Taiji-p (c), LISA, TianQin I, and TianQin II, mainly due to the arm length and orientation of the detector.
In terms of arm length, from Eq.~\ref{Eq:sen_curve1} and Eq.~\ref{Eq:sen_curve2}, it can be seen that the longer the detector arm length results in the better sensitivity. 
Moreover, from TABLE~\ref{Tab:noise}, it can be seen that Taiji's arm length ($3\times10^{9}$ m) is the longest, followed by LISA's arm length ($2.5\times10^{9}$ m), and TianQin's arm length ($\sqrt{3}\times10^{8}$ m) is the shortest, making Taiji detect more resolvable GBs than LISA and TianQin.
In terms of orientation, FIG.~\ref{FIG:response} shows that the detector is most sensitive to signals perpendicular to the constellation plane position ($\theta_d=0^\circ$ or $180^\circ$). 
The density of GBs in the bulge region of the Galaxy is significantly higher than that in the disk region~\cite{Galaxy}, therefore the closer the normal direction of the detector constellation plane is to the Galactic Center ($\lambda = 266.8^\circ,\beta = -5.6^\circ$), the greater the detector response and the more resolvable GBs can be detected.
From FIG.~\ref{FIG:normal_direction}, it can be seen that the normal direction of Taiji-m ($\beta = -30^\circ$ and $\beta = 60^\circ$) is closer to the Galactic Center compared to Taiji-p (c) ($\beta = 30^\circ$ and $\beta = -60^\circ$) over a year, and the normal direction of TianQin I ($\lambda = 120.4^\circ,\beta = -4.7^\circ$ and $\lambda = 300.4^\circ,\beta = 4.7^\circ$) is also closer to the Galactic Center than TianQin II ($\lambda = 30.4^\circ,\beta = 0^\circ$ and $\lambda = 210.4^\circ,\beta = 0^\circ$). Therefore, Taiji-m detects more resolvable GBs than Taiji-p (c), and TianQin I detects more than TianQin II.

For detection on networks, just like the result in a single detector, the arm length and orientation of the detector are the major factors in resolvable GBs detection.
Due to the longer arm length of Taiji and LISA  than that of TianQin, the networks of Taiji and LISA detect more resolvable GBs than individual Taiji or LISA, but the improvement is not significant compared to TianQin's network.
Eq.~\ref{Eq:net_sensitivity} indicates that the reciprocal sensitivity on the network is the sum of the reciprocal sensitivities of each detector. Therefore, as the number of detectors in the network increases, the sensitivity of the network increases, but the increase rate decreases.
In summary, it can be concluded that as the number of detectors on the network increases, the number of resolvable GBs detected will also increase.
The optimal result will be achieved when LISA, Taiji-m, TianQin I and TianQin II are combined as a network.

\subsection{Improvement of sensitivity}
\begin{figure}[ht]
    \begin{minipage}{\columnwidth}
        \centering
        \includegraphics[width=0.9\textwidth,
        trim=0 0 0 0,clip]{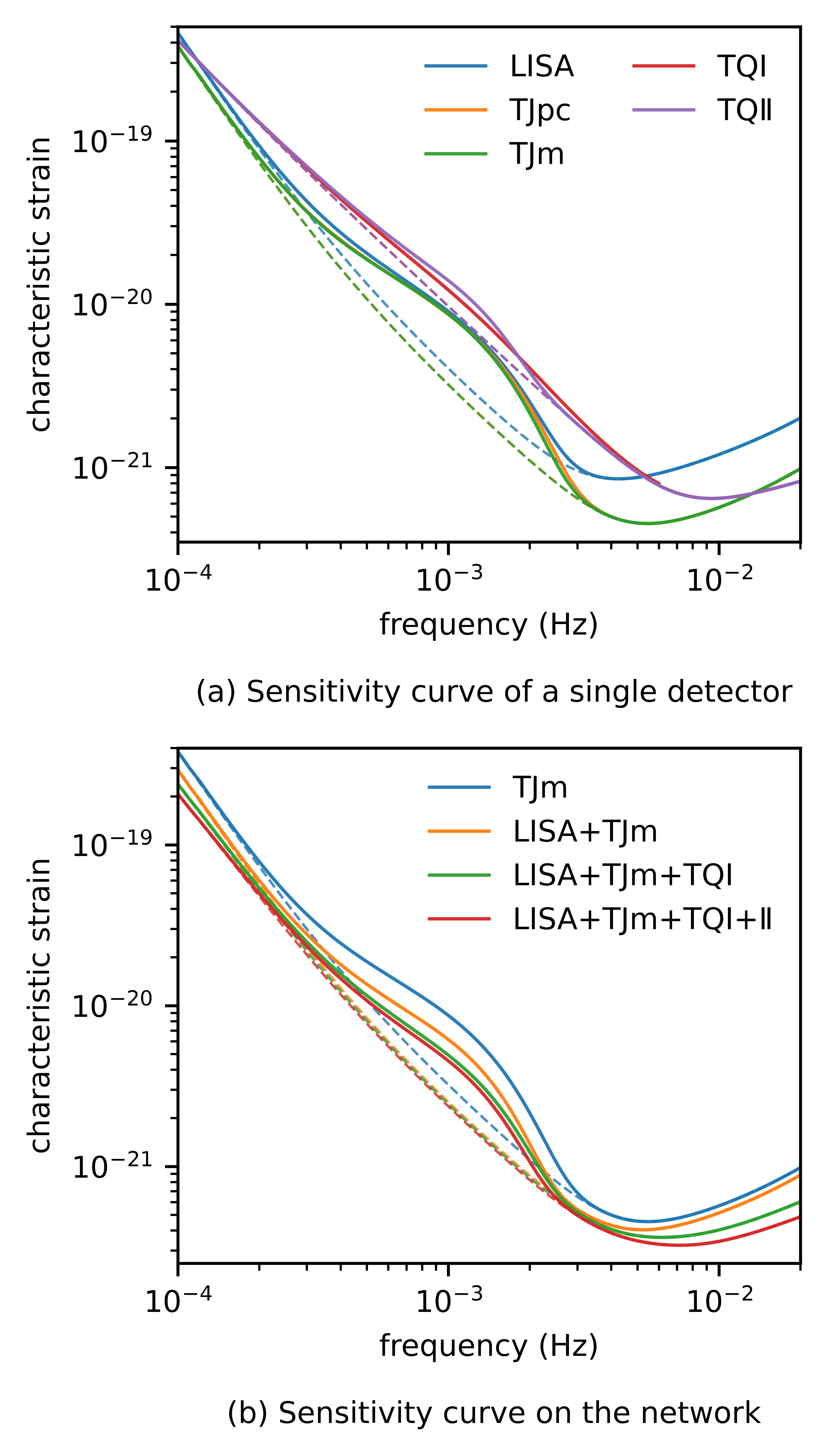}
        \caption{Sensitivity curves of a single detector and detectors on the network for 4 years observation time. The vertical axis adopts dimensionless characteristic strain sensitivity $\sqrt{fS_n(f)}$. The dashed lines are the design sensitivity curves that only considers instrument noise, while the solid lines are the full sensitivity curves that adds confusion foreground.}\label{FIG:sensitivity}
    \end{minipage}
\end{figure}

In order to better show the impact of confusion foreground on the sensitivity curve, and the subtraction of confusion foreground by different number of detectors on the network, we can fit the confusion foreground on logarithmic scale through a polynomial function, which can be written as follows~\cite{TianQinGBs}:
\begin{equation}
    S_c(f) = 10^x
\end{equation}
with
\begin{equation}
    x = \sum_{n = 0}^5{a_n\left[ \log 10\left( \frac{f}{1\text{ mHz}} \right) \right] ^n}
\end{equation}
This fitting is only applicable to the frequency range of 0.1$\sim$6 mHz, and the fitting parameters $a_n$ are listed in TABLE~\ref{Tab:fit}.
Due to different fitting functions, they can affect the final curve. Therefore, the fitting parameters given in TABLE~\ref{Tab:fit} and the curves drawn in FIG.~\ref{FIG:sensitivity} are only used as a reference. In our previous calculations, we estimate confusion foreground using a running median on PSD.

\begin{table*}[ht]
\centering
\renewcommand{\arraystretch}{1.5}
\caption{Fitting parameters of the confusion foreground.}\label{Tab:fit}
\begin{tabular*}{\textwidth}{@{\extracolsep{\fill}}crrrrrr@{}}
\hline
 Detector & $a_0$ & $a_1$ & $a_2$ & $a_3$ & $a_4$ & $a_5$ \\
\hline
LISA & -37.187 & -3.432 & -2.753 & -5.044 & -7.123 & -4.120\\
TJpc & -37.191 & -3.443 & -2.710 &  -4.847 & -6.871 & -4.016 \\
TJm & -37.186 & -3.485 & -3.273 & -5.970 &  -7.926  &-4.785 \\
TQ I & -37.262 & -3.465 & -2.790 &  -2.128 &  1.701 &  2.734 \\
TQ II & -36.999 & -3.177 & -4.288& -11.632& -13.272 & -4.360 \\
LISA+TJm &-37.502 & -3.479 & -3.790 &  -8.853 & -12.191 &  -6.537\\
LISA+TJm+TQ I & -37.739 & -3.472 & -3.384 & -7.864 &-10.962 & -5.755 \\
LISA+TJm+TQ I+II& -37.825 & -3.505 & -4.098& -10.273 &-13.655 & -6.547 \\ 
\hline
\end{tabular*}
\end{table*}

In FIG.~\ref{FIG:sensitivity}(a), we plotted the sensitivity curve of a single detector, and it can be seen that in the part where the confusion foreground affects, the sensitivity curve generated by instrument noise is better in Taiji than in LISA than in TianQin.
In the range of 8$\sim$1.5 mHz, the full sensitivity curves of LISA and Taiji are almost identical, due to the larger response of resolvable GBs in Taiji, resulting in greater confusion foreground.
In the range of 1.5$\sim$3.5 mHz, the full sensitivity of Taiji-m is superior to that of Taiji-p (c), as Taiji-m can detect more resolvable GBs than Taiji-p (c), resulting in lower subtracted confusion foreground.
In the 2$\sim$6 mHz range, the full sensitivity of TainQin I is slightly lower than that of TianQin II, which is also because TainQin I has a greater response to resolvable GBs.

In FIG.~\ref{FIG:sensitivity}(b), we show the sensitivity curves of different numbers of detectors on the network. 
It can be seen that as the number of detectors on the network increases, the sensitivity curve of instrument noise decreases. 
Moreover, because in this range, the sensitivity of TianQin is much lower than that of LISA and Taiji, the sensitivity curve of instrument noise only slightly changes after adding TianQin to the network.
As the number of detectors on the network increases, the more resolvable GBs are the subtracted confusion foreground is smaller, which is sufficient to demonstrate the advantage of detecting on the network for subtracting confusion foreground.

\subsection{SNR and uncertainty}
\begin{figure}[ht]
    \begin{minipage}{\columnwidth}
        \centering
        \includegraphics[width=0.9\textwidth,
        trim=5 5 10 10,clip]{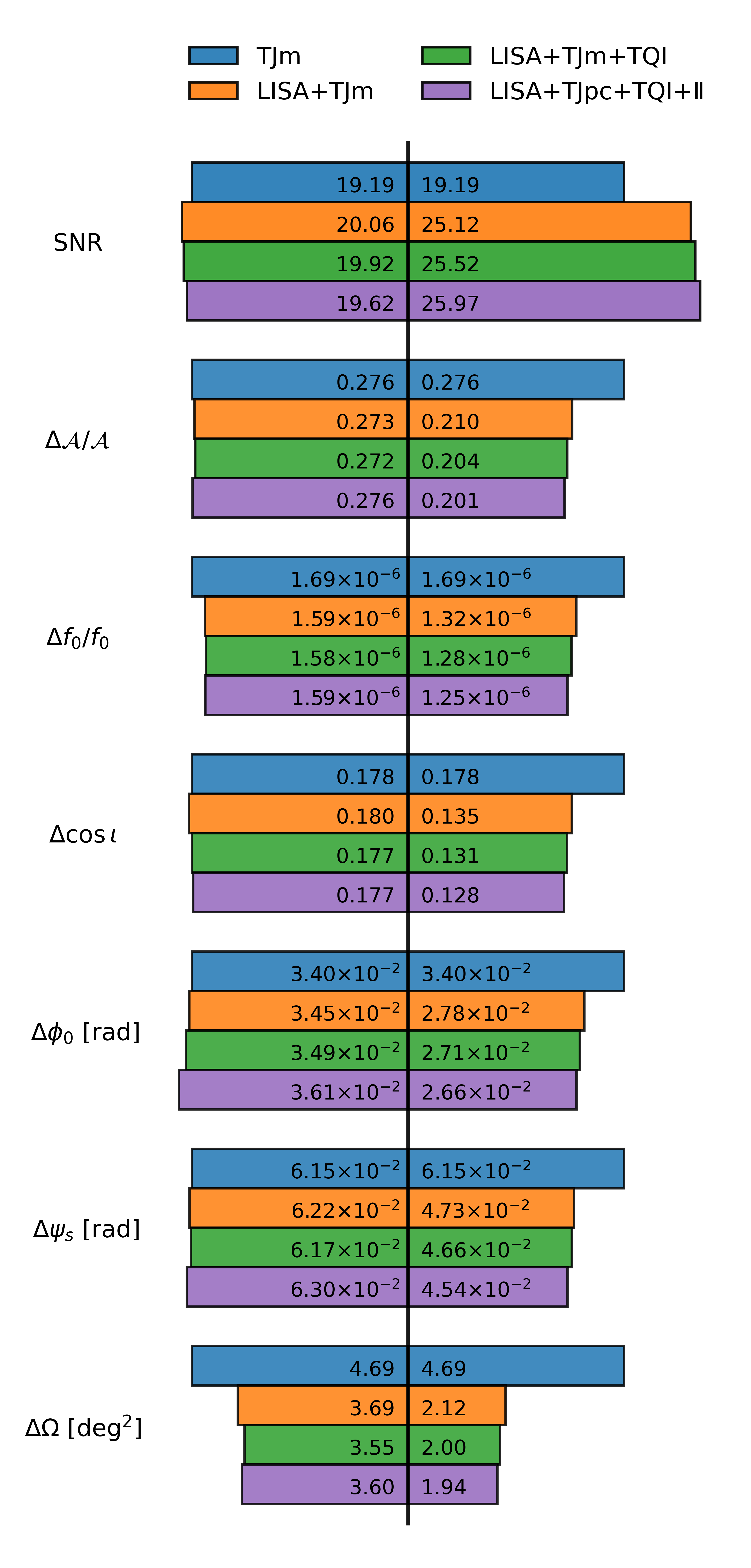}
        \caption{The median SNR and parameter uncertainty on different networks for 4 years observation time. In the left bars we list the SNR and parameter uncertainty on different network of RGB detected by the corresponding detector combination; in the right bars we list the SNR and parameter uncertainty on different network of RGB detected only by Taiji-m. The numbers on the bars are the median of SNR and parameter uncertainty.}\label{FIG:FIM}
    \end{minipage}
\end{figure}

In addition to the number of resolvable GBs detected and the sensitivity curve containing confusion foreground, the uncertainty of parameters for resolvable GBs is also crucial. Therefore, we calculated the FIM on different networks (choosing \textit{TJm}, \textit{LISA+TJm}, \textit{LISA+TJm+TQI} and \textit{LISA+TJm+TQI+II} due to the most number of resolved GBs with 1, 2, 3, 4 detectors respectively)  using Eq.~\ref{Eq:FIM} $\sim$ Eq.~\ref{Eq:net_SNR_FIM} to obtain the uncertainty of different parameters, as shown in FIG.~\ref{FIG:FIM}.

Explanation of result on the right side of FIG.~\ref{FIG:FIM}, for the resolvable GBs detected only by Taiji-m, it can be clearly seen that as the number of detectors on the network increases, the SNR will increase, while the uncertainty of parameters will decrease. 
This is due to the sensitivity improvement for the increased number of detectors on the network.
Similar to the increase rate in the number of resolvable GBs described in Sec.~\ref{sec:Resolvable_GBs}, the magnitude of changes in SNR and uncertainty will decrease as the number of detectors on the network increases.
Increasing from one detector to two has a significant effect, but increasing from two to three is relatively less significant.

Unlike the above situation, in actual detection, the resolvable GB detected by different detector combinations is different.
From the result on the left side of FIG.~\ref{FIG:FIM}, it can be seen that the changes in SNR and uncertainty of resolvable GBs detected on different networks are not as significant as those of the same resolvable GBs. 
Except for the decrease in the uncertainties of GW strain amplitude, frequency, and sky position, there are almost no significant changes in the rest, and even some uncertainties have no decrease but increase. 
For example, the initial phase and polarization angle show a slight increase when the number of detectors on the network increases from three to four.
This is because as the number of detectors on the network increases, the sensitivity improves, making many unresolved GBs become resolvable GBs, adding more low-SNR resolvable GBs. 
Therefore, it is possible that as the number of detectors on the network increases, uncertainty increases instead of decreasing, and SNR decreases instead of increasing.

Nonetheless, as the number of detectors on the network increases, the SNR of the same resolvable GBs increases, and uncertainty decreases. Moreover, after adding more low-SNR resolvable GBs, the overall SNR remains almost unchanged, with some uncertainties significantly decreasing and others slightly increasing, which is sufficient to demonstrate the positive impact of increasing the number of detectors on the network.
Not only these, but also the GW detection of 
resolvable GBs is helpful for the detection of EM bands, constituting Multi-messenger astronomy~\cite{Multi-messenger1,Multi-messenger2,Multi-messenger3,Multi-messenger4}.

The more accurate the GW detection of resolvable GBs parameters, i.e. the lower the uncertainties, the more conducive it is to EM detection.
If the sky position of the source is sufficiently accurate, it is possible to search for EM counterparts through EM follow-up observations
Among all resolvable GBs, the uncertainty of the sky position is less than $1\ \mathrm{ deg}^2$ ($\Delta \Omega<1\ \mathrm{deg}^2$) for 30.2$\sim$31.6\% of resolvable GBs, and less than $0.1\ \mathrm{ deg}^2$ ($\Delta \Omega<0.1\ \mathrm{deg}^2$) for 9.6$\sim$10.3\%.
It can be seen from the data in FIG.~\ref{FIG:FIM} that among all parameters, the frequency measurement of resolvable GBs is the most accurate, of which the uncertainty on $\Delta f_0/f_0$ of 29.2$\sim$32.3\% GB is less than $1\times10^{-6}$ ($\Delta f_0/f_0<1\times10^{-6}$), while the GW frequency $f_0$ is directly related to the period $T_p$ of resolvable GBs ($f_0=2/T_p$), that is, the period can be measured accurately.
Note that as the number of detectors on the network increases, the proportion of the above items will also increase.

On the contrary, the results of EM detection can also serve as a prior to reduce the uncertainty of GW detection.
We adopt the method in Ref.~\cite{TaijiGBs}, which can be used to reduce the uncertainty of parameters from GW data by removing the respective rows and columns in the FIM.
By observing GBs, the inclination angle $\iota $ can be independently determined by EM detection, and we assume that the inclination angle of resolvable GBs can be completely determined. 
By calculating the uncertainty of other parameters through the removed FIM, we found that only the uncertainty on $\Delta \mathcal{A} /\mathcal{A} $ changes significantly, with the mean uncertainty decrease of 91.9$\sim$93.5\% and the median uncertainty decrease of 60.8$\sim$61.9\%.
From Eq.~\ref{Eq:waveform}, there is degeneracy between GW strain amplitude $\mathcal{A}$ and inclination angle $\iota$, which is why determining the inclination angle can significantly improve the measurement of amplitude.
Using the same method, we assume that the EM counterparts can be found through EM detection, that is, the sky position ($\lambda,\beta$) is completely determined. Therefore, the mean uncertainty on $\phi_0$ is reduced by 25.8$\sim$33.6\%, the median uncertainty is reduced by 25.1$\sim$26.9\%, and other parameters will have a decrease of 2$\sim$9\%.
Notice that the above situations are all very idealized and are based on the assumption that a certain parameter of all resolvable GBs is completely determined, which cannot be achieved in practice. Even so, it can also indicate that there is feasibility in reducing the parameter uncertainty of GW detection through EM detection.
In summary, GW detection and EM detection can complement each other, and as the number of detectors on the network increases, the improvement of both will be greater.

\section{Summary and discussion}\label{sce:Summary}
In this paper, we used 1\% of the data in LDC, which is $3\times10^5$ GBs, to simulate the galactic foreground by overlapping GBs as quasi-sinusoidal signals. We treated GB with the SNR greater than 7 as resolvable GBs, studied the number of detected resolvable GBs under different detector combinations and their alternative orbital configurations on the network, calculated the parameter uncertainties of resolvable GBs, and plotted the fitted full sensitivity curve.

Through the iterative method, we predict the number of resolvable GBs detected by different detector combinations on the network. 
In the single detectors, the number of resolvable GBs is arranged in descending order of detected quantity: Taiji-m, Taiji-p (c), LISA, TianQin I, and TianQin II.
The trend of results for different detectors combinations on the network is also similar to that of a single detector.
The optimal combination for each number on the network is \textit{TJm}, \textit{LISA+TJm}, \textit{LISA+TJm+TQI}, and \textit{LISA+TJm+TQI+II}.

Based on the above optimal combinations, we calculate the uncertainty of the parameters of resolvable GBs using FIM. 
As the number of detectors on the network increased, the uncertainty of the same resolvable GBs decreased, and the magnitude of the decrease also decreased.
The uncertainty remained reduced or almost unchanged even when more low-SNR resolvable GBs were detected.
Resolvable GBs with low uncertainty can help EM detection find electromagnetic counterparts and determine the period of GBs, while EM detection can also serve as a prior to reducing the uncertainty of GW detection. 
We find that determining the inclination angle through EM detection can reduce GW strain amplitude uncertainty by $\sim$93\%, and determining the sky position can reduce the phase uncertainty by $\sim$30\%.
Therefore, GW joint detection on the network can complement EM detection, which is conducive to the development of Multi-messenger astronomy.

By fitting the full sensitivity curve containing confusion foreground, it is possible to intuitively see the effect of a single detector and different combinations of detectors on the network on subtracting confusion foreground.
The effect of subtracting confusion foreground is basically proportional to the number of resolvable GBs detected. The more detectors in the network, the better the subtracting effect.

In addition, it should be noted that so far, no space-based GW detector has been launched, so the data related to the space GW detector are simulated and predicted. 
In fact, during the observation, the noise is assumed to be Gaussian and stationary, and the data quality is assumed to be optimal and uninterrupted~\cite{Characterization_LISA}.
We use SNR to define thresholds and distinguish resolvable GBs, which is very useful and efficient to estimate confusion foreground. 
Moreover, we assume that the subtraction of GBs is perfect without residual, which leads to our results being optimal and ideal.
Some new and more practical methods have been proposed, such as iterative subtraction based on Particle swarm optimization algorithm~\cite{network,GBSIEVER2}, search and subtraction using Bayesian evidence ratio~\cite{Bayesian_evidence_ratio}.

In future research, we can delve into multiple aspects to improve our understanding and accuracy of confusion foreground.
Firstly, we can further investigate the relationship between GW detection and EM detection, exploring how to better combine GW detectors and EM detectors to enhance observation and understanding of GBs~\cite{EM_detection}.
Secondly, we can delve deeper into the impact of time-delay interferometr (TDI) technology on confusion foreground, as well as the subtraction of confusion foreground by different generations of TDI and channels~\cite{TDI}
In addition, we can also consider the impact of different population models on confusion foreground to better understand the population distribution and evolution theory of GBs.
Finally, we can also consider the impact of confusion foreground on other GW sources to better evaluate the sensitivity and accuracy of GW detection, and use foreground noise to improve the data processing and analysis methods.
In conclusion, through in-depth research on the above aspects, we can further improve our understanding and accuracy of GW detection, so as to better explore the essence and evolution history of astrophysical events, and provide more valuable data and information for research in Cosmology, Astrophysics and other fields.

\begin{acknowledgements}
This work was supported by the National Key Research and Development Program of China (Grant No. 2021YFC2203004), the National Natural Science Foundation of China (Grant No. 12147102) and Natural Science Foundation of Chongqing (Grant No. CSTB2023NSCQ-MSX0103).
\end{acknowledgements}

\begin{widetext}
\appendix
\section{Coordinate transformation}\label{App:Coordinate_transformation}
The transformation between detector coordinates ($\phi_d$,$\theta_d$) and Ecliptic coordinates ($\lambda$,$\beta$) is based on the method described in Ref.~\cite{response_function}, and the situation in both coordinate frames is shown in FIG.~\ref{FIG:coordinate}.

\begin{figure}[ht]
    \begin{minipage}{\columnwidth}
        \centering
        \includegraphics[width=0.9\textwidth,
        trim=0 0 0 0,clip]{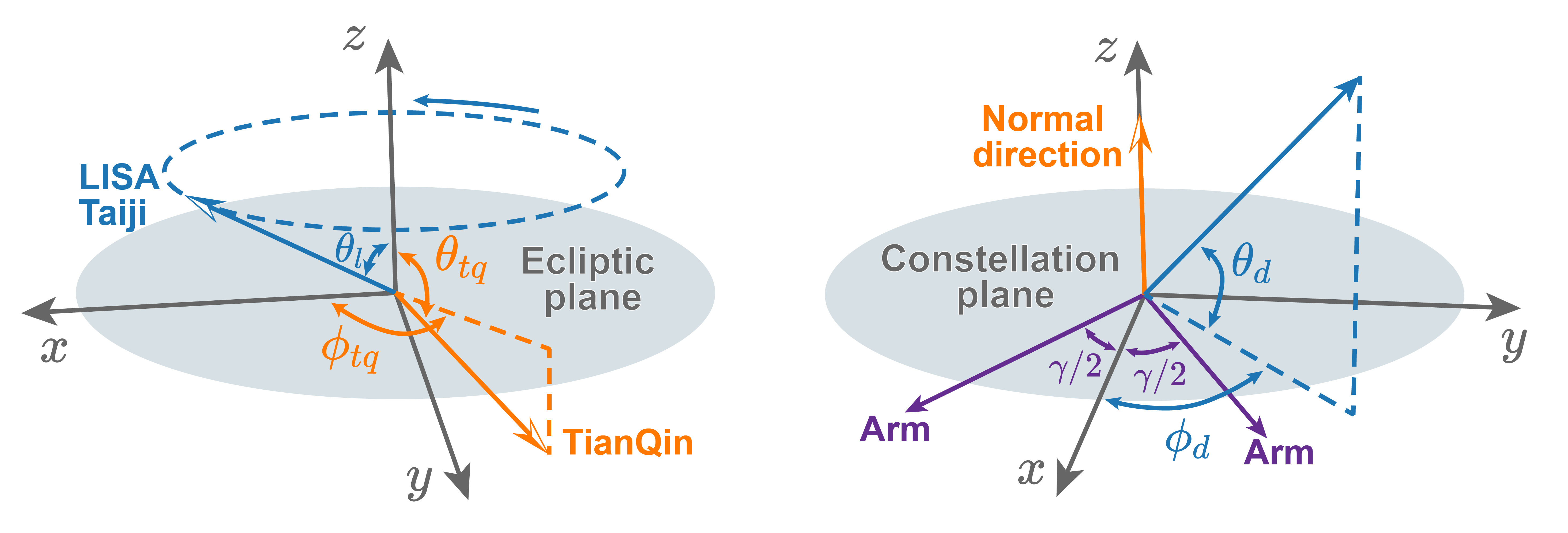}
        \caption{The normal direction of the detector constellation planes in Ecliptic coordinate frame and detector coordinate frame.}\label{FIG:coordinate}
    \end{minipage}
\end{figure}

We can use a rotation matrix $R$ to connect detector coordinates $X^d=\{\sin\theta_d \cos\phi_d,\sin\theta_d \sin\phi_d,\cos\theta_d\}$ and Ecliptic coordinates $X^e=\{\cos\beta \cos\lambda,\cos\beta \sin\lambda,\sin\beta\}$, which can be expressed as:
\begin{equation}
    \begin{aligned}
        X^e&=RX^d\\
        X^d&=R^{-1}X^e
    \end{aligned}
\end{equation}

For LISA and Taiji:
\begin{equation}
    R=
    \left(\begin{array}{ccc}
    \cos \theta_l \cos ^{2} \alpha_{d}+\sin ^{2} \alpha_{d} & (\cos \theta_l-1) \sin \alpha_{d} \cos \alpha_{d} & -\sin \theta_l \cos \alpha_{d} \\
    (\cos \theta_l-1) \sin \alpha_{d} \cos \alpha_{d} & \cos \theta_l \sin ^{2} \alpha_{d}+\cos ^{2} \alpha_{d} & -\sin \theta_l \sin \alpha_{d} \\
    \sin \theta_l \cos \alpha_{d} & \sin \theta_l \sin \alpha_{d} & \cos \theta_l\\
    \end{array}\right)
\end{equation}

For TianQin:
\begin{equation}
    R=
    \left(\begin{array}{ccc}
    \cos \theta_{t q} \cos \phi_{t q} \sin \alpha_{d}+\sin \phi_{t q} \cos \alpha_{d} &\cos \theta_{t q} \cos \phi_{t q} \cos \alpha_{d}-\sin \phi_{t q} \sin \alpha_{d} & \sin \theta_{t q} \cos \phi_{t q} \\
    \cos \theta_{t q} \sin \phi_{t q} \sin \alpha_{d}-\cos \phi_{t q} \cos \alpha_{d} &\cos \theta_{t q} \sin \phi_{t q} \cos \alpha_{d}+\cos \phi_{t q} \sin \alpha_{d} & \sin \theta_{t q} \sin \phi_{t q} \\
    -\sin \theta_{t q} \sin \alpha_{d} & -\sin \theta_{t q} \cos \alpha_{d} & \cos \theta_{t q}
    \end{array}\right)
\end{equation}
where $\alpha_{d}=2\pi f_{sc}t+\frac{2\pi}{3}(n-1)+\alpha_0$, $n$ is the $n$th S/C, $\alpha_0$ is the initial phase, $f_{sc}=1/T_{sc}$ and $T_{sc}$ is the rotation period.
For TianQin, $T_{sc}$=3.65 days and $f_{sc}\simeq3\times10^{-3}$ mHz, but for LISA and Taiji, $T_{sc}$=1 year and $f_{sc}\simeq3\times10^{-6}$ mHz
The angles in the rotation matrix $R$ can be determined from FIG.~\ref{FIG:coordinate}.
For LISA, Taiji-p and Taiji-c, $\theta_l=60^\circ$ and for Taiji-m, $\theta_l=120^\circ$.
For TianQin I, $\theta_{tq}=94.7^\circ,\phi_{tq}=120.4^\circ$ and for TianQin II, $\theta_{tq}=90^\circ,\phi_{tq}=30.4^\circ$.
\end{widetext}
\bibliography{references}
\end{document}